%% file: main.tex
\documentclass{article}

\input{preamble/packages}

\input{preamble/macros}

\graphicspath{ {./images/} }

\title{Smoothing for age-period-cohort models: a comparison between splines and random process}
\author[1*]{Connor Gascoigne}
\author[2]{Theresa Smith}
\author[3]{Andrea Riebler}

\affil[1]{MRC Centre for Environment and Health, Department of Epidemiology and Biostatistics, School of Medicine, Imperial College London, London, UK}
\affil[2]{Department of Mathematical Sciences, University of Bath, Bath, UK}
\affil[3]{Department of Mathematical Sciences, Norwegian University of Science and Technology, Norway}
\affil[*]{Corresponding author. \textit{E-mail address}: c.gascoigne@imperial.ac.uk}

\date{}

\begin{document}

\maketitle

\begin{abstract}
    Age-Period-Cohort (APC) models are well used in the context of modelling health and demographic data to produce smooth estimates of each time trend. When smoothing in the context of APC models, there are two main schools, frequentist using penalised smoothing splines, and Bayesian using random processes with little crossover between them. In this article, we clearly lay out the theoretical link between the two schools, provide examples using simulated and real data to highlight similarities and difference, and help a general APC user understand potentially inaccessible theory from functional analysis. As intuition suggests, both approaches lead to comparable and almost identical in-sample predictions, but random processes within a Bayesian approach might be beneficial for out-of-sample prediction as the sources of uncertainty are captured in a more complete way.
\end{abstract}

\textbf{Key words:} Age-period-cohort, identifiability, smoothing, penalised splines, random processes, Bayesian model

\input{documents/manuscript}

\clearpage

\bibliographystyle{myunsrtnat}
\bibliography{preamble/library}

\end{document}

%% file: preamble/packages.tex
\usepackage[left = 2cm, right = 2cm, top = 2cm, bottom = 2cm]{geometry}

\usepackage[comma, sort&compress]{natbib}
\defcitealias{KDHS2014}{KDHS, 2015} % shorthand in text citation
\usepackage[utf8]{inputenc}
\usepackage{authblk} % \author as a footnote
\usepackage{setspace} % spacing between lines i.e. \doublespace
\usepackage[english]{babel} % special characters
\usepackage{mathpazo} % Palatino font
\usepackage{amsmath, amsthm, mathtools, amsfonts} % maths related notation
\usepackage{lineno} % adds line number to selected paragraphs
\usepackage[table]{xcolor} % to include coloured text and in cell entry
\usepackage{longtable, lscape} % have tables over multiple pages
\usepackage[colorlinks=true,allcolors={blue!80!black}]{hyperref} % blue for online
\usepackage[capitalise, noabbrev]{cleveref} % LOAD AFTER HYPERREF :: automatically detects if table, figure, etc... 
\usepackage{multirow} % to group multiple rows in tables
\usepackage[title,titletoc]{appendix} % for appendix
\usepackage{tikz}
\usetikzlibrary{matrix, calc}

\usepackage{caption} % for sub figures
\usepackage{subcaption} % for sub figures
\usepackage{graphicx} % including graphics
\usepackage{rotating} % sideways figures and tables

%% file: preamble/macros.tex
\makeatletter
\def\hlinewd#1{%
  \noalign{\ifnum0=`}\fi\hrule \@height #1 \futurelet
   \reserved@a\@xhline}
\makeatother
\newcommand{\mgcv}{\texttt{mgcv}}
\newcommand{\inla}{\texttt{R-INLA}}

\newcommand{\ageFull}{f_A}
\newcommand{\perFull}{f_P}
\newcommand{\cohFull}{f_C}

\newcommand{\ageFullB}{\bs{f}_A}

\DeclareMathOperator*{\argmax}{arg\,max} % arg max
\DeclarePairedDelimiter\abs{\lvert}{\rvert} % absolute
\newcommand{\pa}[1]{\left(#1\right)} % stretchable parentheses
\newcommand{\bra}[1]{\left[#1\right]} % stretchable square brackets
\newcommand{\bs}[1]{\boldsymbol{#1}} % bold symbol
\newcommand{\EE}{\mathbb{E}}

\newcommand{\II}{\mathbb{I}}

\newcommand{\mL}{\mathcal{L}}

%% file: documents/manuscript.tex
% \textbf{Journal: Sociological Methods \& Research}

\section{Introduction}

Two important goals for any researcher interested in modelling how disease and demographic rates vary over time are (1) validating hypotheses about what is driving the underlying phenomena of interest and (2) forecasting the evolution of rates into the future. Both of these goals work towards the ever-important data-driven approach to effective and efficient policy evaluation and recommendation.

The class of so-called Age-Period-Cohort (APC) models is a popular tool for modelling the evolution of rates over time when there are different patterns of change by age group. In APC models, we consider three time scales: age—the age of an individual when the event of interest occurs; period—the time (often a year) that the event occurs; and cohort—the time (usually a year) that the individual was born. Applications of APC models can be found in a range of health contexts, such as modelling prostate cancer, thyroid cancer, or stomach cancer \citep{li2020long, liu2019, papoila2014stomach}, and sociological contexts, such as exploring trends in suicide or opioid deaths \citep{kim2021age, chernyavskiy2020spatially}.

Although popular, users of APC models must overcome a key technical issue called the 'APC identification problem.' Because the three time scales of interest are intrinsically connected (e.g., given the birth year (cohort) and age at the event of an individual, we can calculate the year of the event (period) as cohort + age), we cannot estimate linear trends along all three time scales simultaneously. A common solution to the identification problem is to parameterise the temporal trends into identifiable quantities that can be fully estimated.

On top of the identification problem, additional identification issues may arise (in the form of artificial cyclic patterns in the estimates) when the groups for age, period, and cohort are aggregated in non-equal interval widths (i.e., groups that do not contain the same number of years). In these scenarios, it has been shown that smoothing in APC models is an effective method to relieve the additional identification problems \citep{holford2006approaches, riebler2010analysis}. However, \citet{gascoigne2023penalized} has shown that the effectiveness of the smoothing function is dependent upon its specification, and for estimates that alleviate the additional identification problem regardless of the parameterisation, a penalty on the second derivative (a measure of how 'wiggly' the function is) is essential.

For smoothing in a frequentist setting, splines are often used, while random processes (i.e., random walk models) are used in a Bayesian setting. For smoothing with a penalty on the second derivative, (cubic) penalised smoothing splines and Random Walk 2 (RW2) random processes are standard approaches used in frequentist and Bayesian settings, respectively.

Forecasting, or predicting, is important for policy planning such as the allocation of public health funds. When predicting, there is often no single best temporal scale to use for prediction. As APC models incorporate three influential time scales, they are often used when predictions are needed \citep{berzuini1994bayesian}. In the context of APC models, suitable predictions are made from models where the temporal terms are correctly identified \citep{kuang2008forecasting, smith2016review}. Consequently, smooth estimates of the age, period, and cohort trends are vital to ensure adequate and suitable predictions can be made from APC models.

The choice of implementing APC models in a frequentist or Bayesian setting is often based on philosophical reasons or the background of the applied scientist. In this article, we compare both approaches based on in-sample and out-of-sample predictions to highlight important differences in how sources of uncertainty are captured.

The rest of the article is as follows. In Section 2, we present the alcohol and self-harm deaths data used as a real data illustration. In Section 3, we present the APC model. In Section 4, we give an explanation of smoothing approaches using penalised splines and random processes and highlight the theoretical link between them. In Sections 5, we outline a simulation study used to show the similarities and differences between the two approaches when using the methods practically. In Section 6, we show the results of the methods applied to the alcohol and self-harm death data. Finally, in Section 7, we finish with a discussion.

%\section{Age-Period-Cohort model} \label{Section: APC}
\section{Alcohol and self-harmed deaths in England and Wales, 2006-2021} \label{Section: data}

The World Health Organization (WHO) has identified increasing mental health awareness as a key target of theirs to achieve their sustainable development goals (SDGs) 3.4 \citep{who2022sustainable}, due to significant associations between mental health disorders and non-communicable diseases \citep{stein2019integrating}. With over 700,000 global deaths from suicide every year, suicide is the fourth leading cause of deaths among those aged 15 -- 29 \citep{who2019suicide} and is a key indicator for mental illness. APC models are often considered in the context of modelling suicides to explore differences in suicide rates due to age and short-term (period) and long-term (cohort) time frames. Smoothing functions for APC models have been used to model suicides in Brazil \citep{rodrigues2023influence}, China \citep{wang2016temporal}, Hong Kong and Taiwan \citep{chen2021age}, Korea \citep{park2016age}, and Switzerland \citep{riebler2012gender} amongst others.

\sloppy In this article, we use mortality attributable to suicide in England and Wales to compare frequentist or Bayesian smoothing methods in the context of APC models. From the UK's Office for National Statistics (ONS), we downloaded counts of suicide using the International Classification of Diseases version 10 (ICD-10) codes X60 -- X84 (intentional self-harm). The counts were given in yearly periods from 2006 to 2021 and in five-year age bands between the ages of 25 to 85. The data from 2013 onwards was extracted through the ONS Nomis tool (https://www.nomisweb.co.uk/) and data from 2006 to 2012 was assembled from individual annual reports. In addition, mid-year population estimates were extracted by five-year age groups in all years from the ONS Nomis tool.

In the suicide-related deaths data, there are, on average, 299.13 events per age-year combination. To explore whether the magnitude of the observation has an effect on the performance of frequentist or Bayesian smoothing methods, we included a second dataset of deaths due to mental and behavioural disorders due to the use of alcohol (ICD-10 code F10), for which the average number of events per age-year combination is 46.63. Alcohol abuse (and subsequent death) has a strong comorbidity with mental illnesses such as anxiety and depression \citep{appleton2018association}. In the following, we refer to these two datasets as alcohol-related and self-harm-related deaths.

\begin{figure}[!h]
    \centering
    \begin{subfigure}{.5\textwidth}
        \centering
        \includegraphics[width=\linewidth]{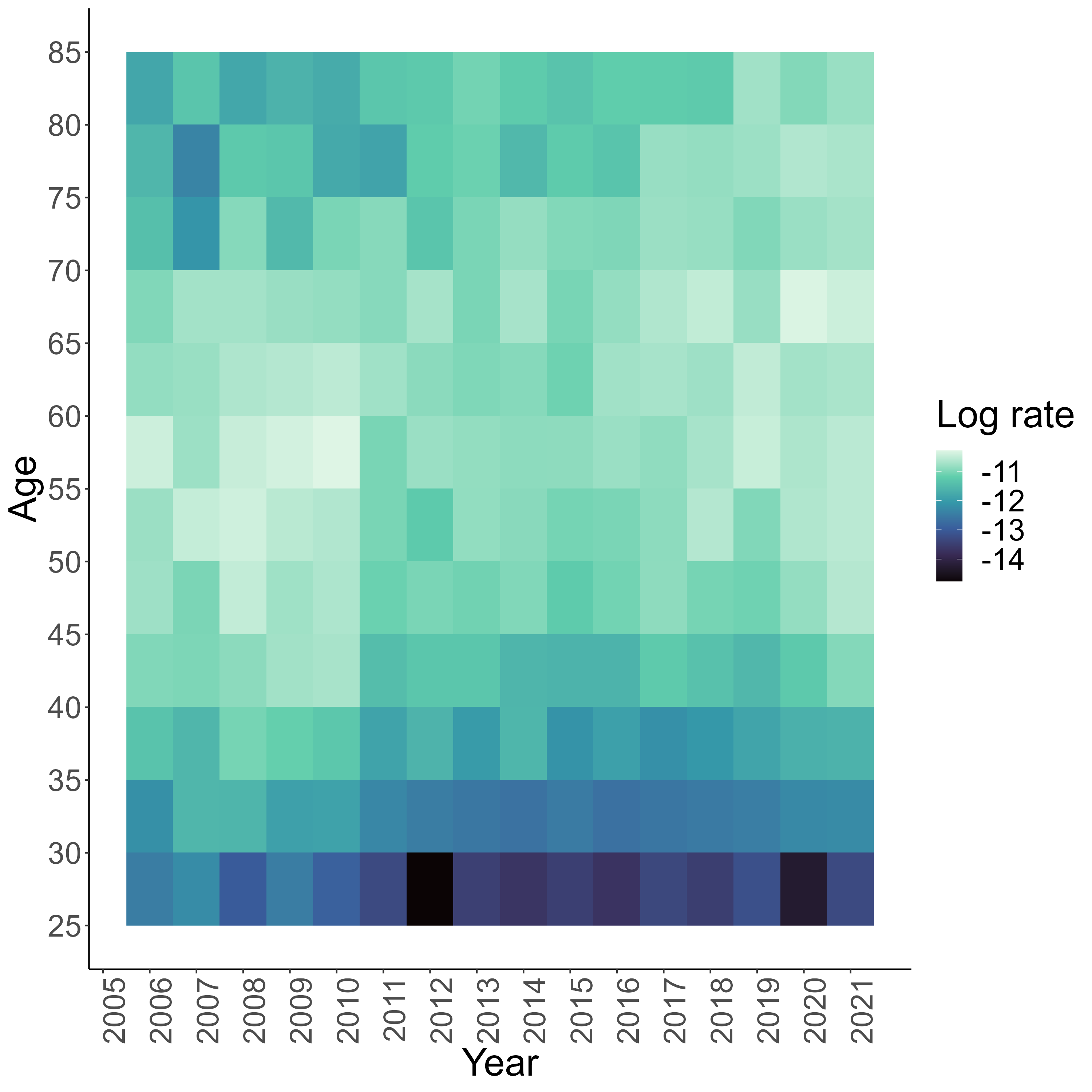}
        \caption{Alcohol related deaths}
        \label{Fig: alcoholObservedHeatmap_25plus}
    \end{subfigure}%
    \begin{subfigure}{.5\textwidth}
        \centering
        \includegraphics[width=\linewidth]{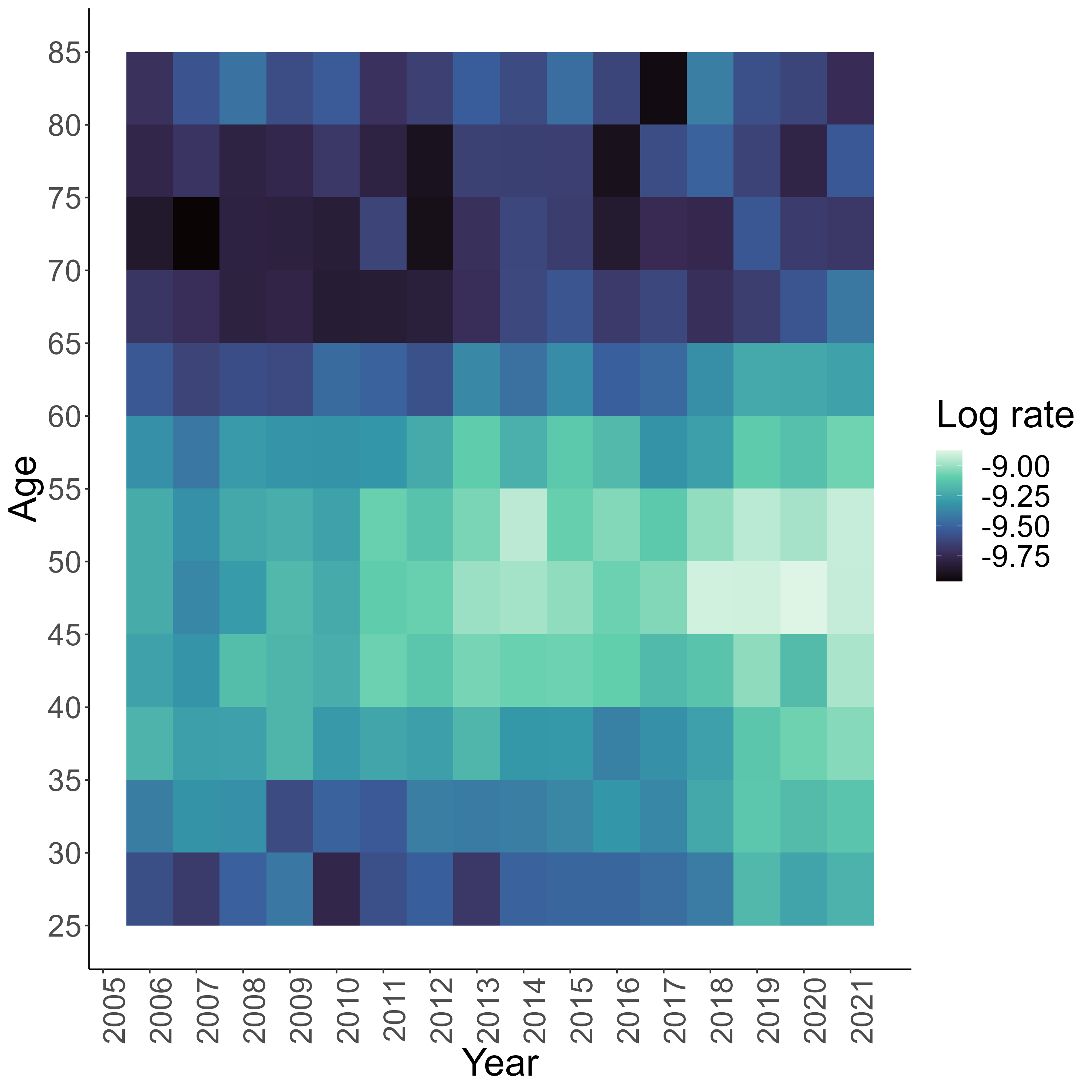}
        \caption{Self-harm related deaths}
        \label{Fig: selfHarmObservedHeatmap_25plus}
    \end{subfigure}
    \caption{Deaths due to alcohol (a) and self harm (b) for the years 2006--2021 and ages 24--84. Period is grouped into single years and operates along the $x$-axis. Age is grouped into five-year ages groups and operates along the $y-axis$. Cohort operates along the $y = x$ axis. Deaths are reported as log-rates with the dark-to-light colouring indicating lower-to-higher rates.}
    \label{Fig: ObservedHeatmap_25plus}
\end{figure}

%https://www.ons.gov.uk/peoplepopulationandcommunity/birthsdeathsandmarriages/deaths/datasets/deathsregisteredinenglandandwalesseriesdrreferencetables

Figure \ref{Fig: ObservedHeatmap_25plus} shows the age-specific log-rates over time for both alcohol- and self-harm-related deaths. Using the natural logarithmic scale allows for a closer inspection of the trends in age, period, and cohort. An extra $1/2$ event was added to each mortality count to avoid taking logs of $0$. For both causes, there is substantial variation in risk according to age (looking along the $y$-axis). For example, the risk of alcohol-related deaths is much lower for the younger age groups (25 -- 30 and 30 -- 35) than in any other age group, and the risk of death related to self-harm is lowest in the youngest (25 -- 30) and the older (60+) age groups, with the ages in between having much higher rates.

There is some evidence of trends over the year of death, but they are small in comparison to trends in age of death. Finally, there are also age-by-year interactions that could indicate cohort effects. For example, ages 70+ had a lower risk of alcohol-related deaths in the years 2006 -- 2010 than in more recent years, and ages 25 -- 30 had a lower risk of death relating to self-harm in the years 2006 -- 2013 than in 2013 and onwards.

\section{Age-Period-Cohort model} \label{Section: APC}

A traditional APC model to capture fluctuations in expected mortality or incidence rates is a log-linear model containing additive functions of age, period, and cohort. Due to the identification problem (cohort = period - age), the temporal linear trends are unidentifiable, and APC models are often reparameterised into identifiable quantities. A common method of reparameterization was proposed by \citet{holford1983estimation}, where the temporal terms were partitioned into a linear trend (slope) and their respective (orthogonal) functions of curvature, with the latter being identifiable.

Following a rare-event approach, we assume the number of events $y_{ap}$ follows a Poisson distribution: $y_{ap} \mid \eta_{ap} \sim \text{Poisson}(N_{ap} \exp(\eta_{ap}))$ for age $a = 1, \dots, I$ and period $p = 1, \dots, J$, with $N_{ap}$ denoting the population size. Then, an APC model reparameterized into linear trends and identifiable functions of curvature is:
\begin{equation}
    \label{Eq: APC model}
    \eta_{ap} = \beta_0 + \beta_1{s_{a}} + \beta_2{s_{p}} + \ageFull\pa{a} + \perFull\pa{p} + \cohFull\pa{c}
\end{equation}
 where $\eta_{ap} = \log\pa{\EE\bra{y_{ap}/N_{ap}}}$ is the log of the expected rate, $\ageFull, \perFull,$ and $\cohFull$ are the age, period, and cohort functions of curvature, and $c = 1, \dots, K = R \times \pa{I - a} + p$ is the index for the cohort, with $R$ as the ratio between the number of years in the age group to the number of years in the period group.%, so that for the datasets described in Secton \ref{Section: data} we set $M=5$.

In any reparameterisation of an APC model, an arbitrary choice has to be made. In Equation \ref{Eq: APC model}, that choice is which two (out of the three) slopes to include (i.e., $s_{a}, s_{p}$ or $s_{c}$ for age, period, and cohort). \citet{holford1983estimation} showed that the choice of what slope to include did not affect curvature estimates, and by selecting two of the three slopes, it implicitly assumes the effect of the dropped slope is zero, and the interpretation of the remaining slopes is the trend in that slope, plus something from the dropped slope. We dropped the cohort slope. Therefore, we implicitly assumed $s_c = 0$, and $s_a$ and $s_p$ are the linear trends in age + cohort and period + cohort, respectively.

For the data on deaths related to alcohol and self-harm, we have $J = 16$ periods and $I = 12$ age groups. Data is grouped into unequal intervals with the period being in a single year and age being in five-year interval widths; therefore, $R = 5$ when indexing cohort. When APC models are fit to data aggregated in such a way, the (previously identifiable) curvature terms of the period and cohort are no longer identifiable and display an artificial cyclic pattern \citep{holford2006approaches, riebler2010analysis, held2012conditional}. Modelling the curvature functions using smoothing functions has been used as an effective solution to this additional form of identifiability issues, but this depends upon arbitrary choices in how to define the choice of smoothing function. It has been shown the estimates can be robust to both the additional identification problem and the specification of the smoothing function when including a penalty term on the second derivative of the estimate of the smoothing function \citep{gascoigne2023penalized}.

\section{Smoothing approaches}

We now describe the theoretical parallels between the smoothing approaches of penalised splines and random processes. The connection has been discussed previously in the context of ecology and the use of stochastic partial differential equations \citep{miller2020understanding}. However, we consider the connection in relation to health and demographic modelling and are using random walk models within the context of an APC framework.

For the purpose of explanation, we describe smoothing for a simple univariate function and consider the simple one-dimensional smooth over age: 
\begin{equation*}
    \eta_{a} = \beta_0 + \beta_1 s_a + \ageFull\pa{a}.
\end{equation*}

\subsection{Smoothing splines}\label{SubSection: splines}

In a frequentist paradigm, an estimator of $\ageFull$ can be found by maximising the following penalised log-likelihood,
\begin{equation}
    \label{Eq: Cubic Penalised Log-Likelihood}
    \widehat{\beta}, \widehat{\ageFull}, \widehat{\theta} = \argmax_{\beta, \ageFull, \theta} \left[l\pa{\bs{\beta}, \ageFullB, \bs{\theta}} +  \lambda \int \ageFull''\pa{a}^2 d{a}\right]
\end{equation}
where $l\pa{\bs{\beta}, \ageFullB, \bs{\theta}} = \log\mL\pa{\bs{\beta}, \ageFullB,\bs{\theta}}$ is the log-likelihood, and  $\lambda \int \ageFull''\pa{a}^2 d{a}$ is a penalty function on the second derivative of the smooth function $\ageFull$ with smoothing parameter $\lambda$ that controls the trade-off between model fit and smoothness. The inclusion of the penalty function on the second derivative of $\ageFull$ penalises $\ageFull$ when it deviates from linearity. If $\lambda = 0$, there is no cost for fitting complicated functions and $\widehat{\ageFull}$ can be extremely `wiggly'. As $\lambda \rightarrow \infty$, the cost for fitting a complicated function increases and $\widehat{\ageFull}$ is forced to be closer to a simple polynomial. 

To make maximising Equation \ref{Eq: Cubic Penalised Log-Likelihood} tractable, we use a finite basis approximation to true function $\ageFull$. Within the context of APC modelling, smoothing splines have been used to approximate the true function $\ageFull$ on several occasions \citep{heuer1997modeling, holford2006approaches, carstensen2007age, fu2008, jiang2014}. A spline basis is a set of polynomial (basis) functions which are based on points called knots. Given $g_t\pa{x}$, the $t^\text{th}$ basis function, $f$ is approximated with a spline as follows
\begin{equation*}
    \label{Eq: Basis Approximation For f}
    \ageFull\pa{a} = \sum_{t = 1}^{T} g_{t}\pa{a} \gamma_{t} = \bs{Z}\bs{\gamma}
\end{equation*}
where $T$ is the number of basis function, $\gamma_{t}$ are the unknown weights to be estimated and $\bs{Z}$ is an $n \times T$ matrix of basis vectors. We consider three examples of spline basis functions, which are Thin Plate Regression Splines (TPRS), Cubic Regression Splines (CRS) and B-Splines (BS). Whilst all are suitable for our application, they have their advantages and disadvantages. TPRS can smooth in multiple dimensions and do not necessarily need the number of knots specified, but they are computationally expensive. CRS are computationally cheaper than TPRS, but only smooth in one dimension. BS are sparse and can be flexibly paired with penalties of different orders, however this causes the interpretation of the penalty to be less clear when compared to derivative-based. Figure S1 in the Supplementary Material shows examples of a CRS, BS and TPRS bases defined by five knots. For a detailed description of these spline bases and other, see \citet[Chapter 5]{wood2006generalized}.

With a given basis representation, $\bs{g}$, with unknown weights, $\bs{\gamma}$, the penalty function for $\ageFull$ can be rewritten
\begin{equation*}
    \label{Eq: Basis Approximation For Cubic Penalty}
    \int \ageFull''\pa{a}^2 d{a} = \bs{\gamma} \int \bs{g}^T\pa{a}\bs{g}\pa{a} d{a} \bs{\gamma} = \bs{\gamma} \bs{S} \bs{\gamma}
\end{equation*}
where $\bs{S} = \int \bs{g}^T\pa{a}\bs{g}\pa{a} d{a}$ is known as the penalty matrix. Therefore, the penalised log-likelihood to be maximised can be re-written in terms of the finite basis approximation to each of the smooth functions,
\begin{equation}
    \label{Eq: Basis Approximation For Cubic Penalised Log-Likelihood}
    l_p\pa{\bs{\beta}, \bs{\gamma}, \bs{\theta}} = l\pa{\bs{\beta}, \bs{\gamma}, \bs{\theta}} + \lambda \bs{\gamma} \bs{S} \bs{\gamma}.
\end{equation}
where $l_p\pa{\bs{\beta}, \bs{\gamma}, \bs{\theta}} = \log\mL_p\pa{\bs{\beta}, \bs{\gamma}, \bs{\theta}}$ is the penalised log-likelihood that is maximised by $\widehat{\bs{\beta}}$, $\widehat{\bs{\gamma}}$ and $\widehat{\bs{\theta}}$, where $\bs{\theta}$ is a vector of any other parameters in the likelihood (e.g., the dispersion parameter in a negative binomial model). 

\begin{figure}
    \centering
    \begin{subfigure}[b]{.48\textwidth}
        \centering
        \includegraphics[width=\linewidth]{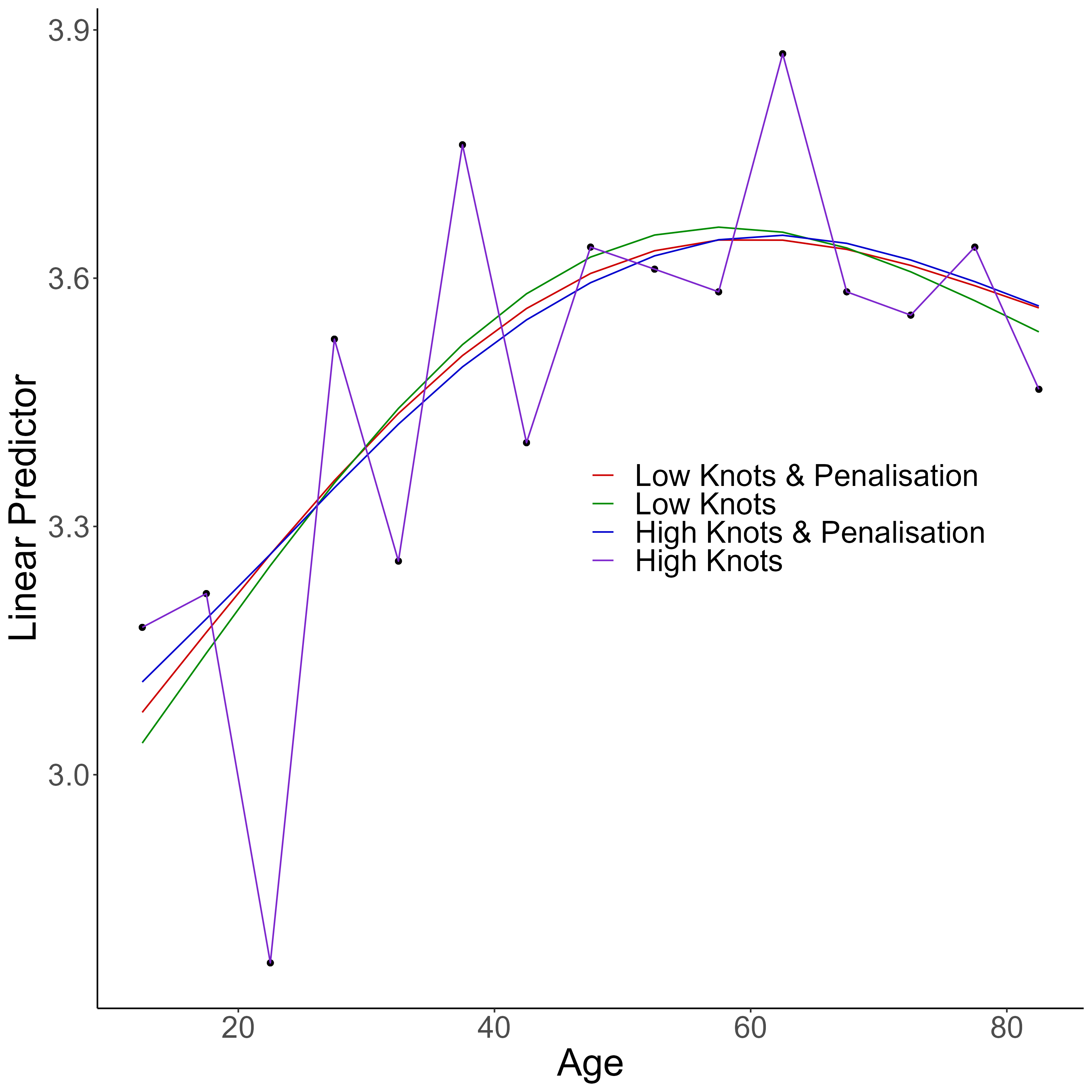}
        \caption{Importance of penalisation.}
        \label{Fig: whyIncludePenalty}
    \end{subfigure}%
    \quad
    \begin{subfigure}[b]{.48\textwidth}
        \centering
        \includegraphics[width=\linewidth]{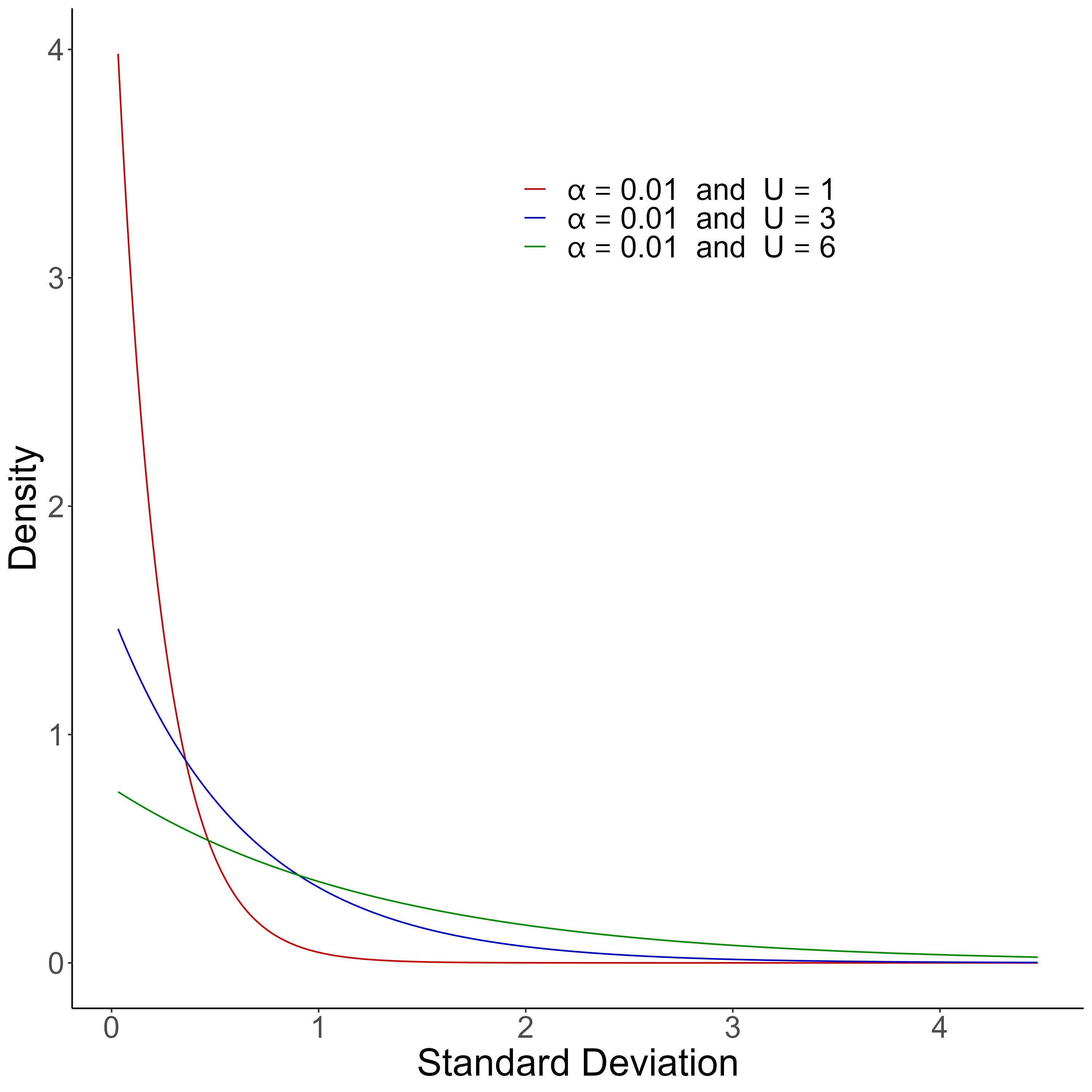}        
        \caption{Importance of prior specification.}
        \label{Fig: pcPriorsPlot}
    \end{subfigure}
    \caption{The left hand plot (a) shows the Importance of penalisation (see section \ref{SubSection: splines}) to reducing overfitting. Four thin plate regression spline models with or without penalty and with large or small number of knots are fit to toy data (black dots). The right hand plot (b) shows the importance of prior specification (see section \ref{SubSection: RW2}). Three difference PC priors specifications of the standard deviation $\sigma = \sqrt{1/\tau}$ are compared with for different choices of $U$ in $\text{P}(\sigma >U) = \alpha$, with $\alpha=0.01$}
    \label{fig:test}
\end{figure}

Where pre-specification of knots is required, an important aspect is selecting the number of knots to use and where to place them. The more knots, the more flexible the spline is, but without a penalty, this can lead to overfitting. Using a toy dataset, Figure \ref{Fig: whyIncludePenalty} highlights how a penalty reduces the importance of the selection of the number of knots. In the toy dataset, the number of unique ages was $A = 15$. We used 3 and 15 knots to define the low and high knot choices, respectively. High knots without a penalty clearly overfit. High knots with a penalty form a smooth curve similar to that of low knots. The inclusion of the penalty ensures there is no overfitting, while maximising the amount of information to be gained by including more knots.

\subsection{Random processes} \label{SubSection: RW2}

In a Bayesian paradigm, random processes are commonly used for smoothing. For temporal smoothing in sociological and health contexts, random walks of order 1 and 2 (RW1 and RW2, respectively) are very popular, so that they are also widely used within APC models for both estimation and prediction \citep{berzuini1993bayesian, berzuini1994bayesian, besag1995bayesian,  knorr2001projections, riebler2010analysis, riebler2012estimation, riebler2016intuitive, cameron2021}. Recently, \citet{okui2021} used random walk priors in an APC model to analyse the prevalence of common psychiatric disorders in Japan. RW1 and RW2 models achieve smoothing by penalising deviations from a constant or linear trend, respectively, and as they are (intrinsic) Gaussian Markov random fields (GMRFs) with sparse inverse covariance matrices, i.e.\,precision matrices, they offer good computational properties. 

Similarly to the penalised smoothing splines, a RW2 penalises deviations in linearity \citep{rue2005gaussian}. Assuming $\ageFullB$ follows a RW2 model, the second differences have the distribution
\begin{equation*}
    \label{Eq: Second Difference Distribution for RW2}
    \Delta^2 {\ageFull}(a) \sim_\text{iid} \text{N}\pa{0, \tau^{-1}},\, a < I-2
\end{equation*}
where there is a flat prior on first two time points and $\tau$ is the precision (inverse variance) parameter. The precision $\tau$ controls the trade-off between smoothness and closeness to the data (it is the smoothing parameter), and has parallels with the smoothing parameter $\lambda$ in the splines framework. For example, as $\tau \rightarrow \infty$, the distribution of $\ageFull$ shrinks towards a straight line. The joint density of $\ageFullB$ is defined, 
\begin{equation}
    \label{Eq: Joint Density of the RW2}
    \begin{split}
        \pi\pa{\ageFullB | \tau} & \propto \tau^{{\pa{I - 2}}/{2}} \exp \pa{-\frac{\tau}{2} \sum_{a = 1}^{I - 2} \bra{ {\ageFull}(a) - 2{\ageFull}(a+1) + {\ageFull}(a+2) }^2 } \\
        & = \tau^{{\pa{I - 2}}/{2}} \exp \pa{ \frac{1}{2} \ageFullB^T \bs{Q} \ageFullB }
    \end{split}
\end{equation}
with precision matrix
\begin{equation*}
    \label{Eq: RW2 Precision}
    \bs{Q} = \tau \bs{R} = \tau
    \begin{pmatrix*}[r]
        1 & -2 & 1 \\
        -2 & 5 & -4 & 1 \\
        1 & -4 & 6 & -4 & 1 \\
        & \ddots & \ddots & \ddots & \ddots & \ddots \\
        & & 1 & -4 & 6 & -4 & 1 \\
        & & & 1 & -4 & 5 & -2 \\
        & & & & 1 & -2 & 1 \\
    \end{pmatrix*}
\end{equation*}
where $\bs{R}$ is referred to as the structure matrix of rank $I-2$. Due to this, the RW2 precision matrix, $\bs{Q} = \tau \bs{R}$, is rank deficient, so that the RW2 is an improper or intrinsic GMRF. Of note, the definition of the RW1 is based on the first differences and analogous to the definition of the the RW2, for details we refer to \citet[Section 3.3.1]{rue2005gaussian}.

Within a Bayesian framework a prior distribution needs to be put on the smoothing parameter $\tau$. As recommended by \citet{simpson2017penalising} we use penalised complexity (PC) prior for $\tau$, which is a type-2 Gumbel distribution for $\tau$ or equivalently an exponential distribution on the standard deviation $\sigma= \sqrt{1/\tau}$ with parameter $\kappa$. The rate parameter $\kappa$ is chosen based on a probability contrast for $\sigma$, specifying $\text{Prob}(\sigma >U) = \alpha$, with $U>0$ and $\alpha \in (0,1)$, such that $\kappa = -\ln(\alpha)/U$. For details, we refer to \citet{simpson2017penalising}. Figure \ref{Fig: pcPriorsPlot} shows the three PC priors we will consider in the following, using $\alpha=0.01$ and either $U=1$, $U=3$ or $U = 6$.

\subsection{Connections between penalised smoothing splines and random processes}

There are theoretical parallels between estimates produced from penalised smoothing spline models and random walk models based on theory from functional analysis. We wish to identify the key information from this theory so that a general practitioner can make the connection and use the methods interchangeably. When estimating $\ageFull$, we have shown above that we can either define a finite-dimensional approximation to $\ageFull$ and then estimate the parameters of the approximation by maximising the penalised log-likelihood, or we can place a prior on $\ageFull$ and define estimates of the model parameters by evaluating the posterior distribution via Bayes rule.

To see the relation between a penalised smoothing spline and RW2 prior model, lets rewrite the penalised log-likelihood in Equation \eqref{Eq: Basis Approximation For Cubic Penalised Log-Likelihood} in terms of a likelihood,
\begin{equation*}
    \label{Eq: Cubic Penalised Likelihood}
    \mL_p\pa{\bs{\beta}, \bs{\gamma}, \bs{\theta}} = \mL\pa{\bs{\beta}, \bs{\gamma}, \bs{\theta}} \times \exp\pa{- \lambda \bs{\gamma}^T \bs{S} \bs{\gamma}}.
\end{equation*}
If we consider the key concept in Bayesian inference ($posterior \propto prior \times likelihood$), $\mL_p\pa{\bs{\beta}, \bs{\gamma}, \bs{\theta}}$ and $\mL\pa{\bs{\beta}, \bs{\gamma}, \bs{\theta}}$ are equivalent to the posterior distribution and the likelihood function, respectively and $\exp\pa{- \lambda \bs{\gamma}^T \bs{S} \bs{\gamma}}$ can be thought of as the prior distribution of the parameters $\bs{\gamma}$. The prior distribution
\begin{equation*}
    \label{Eq: Prior Distribution From Penalised Likelihood}
    p\pa{\bs{\gamma} | \lambda} \propto \exp\pa{- \bs{\gamma}^T \bs{Q}_\lambda \bs{\gamma}}
\end{equation*}
where $\bs{Q}_\lambda = \lambda \bs{S}$ is of the form of an improper GRF prior on $\bs{\gamma}$ with mean zero and precision $\bs{Q}_\lambda$, i.e., $\bs{\gamma} \sim \text{N}\pa{\bs{0}, \bs{Q}_\lambda^{-1}}$, \citep{wood2006generalized, yue2012priors}. The GRF is improper as it is rank deficient by the size of the null space of the penalty matrix, $\bs{S}$. In our penalised spline, the dimension of the null space is two, therefore the precision matrix is of an improper GRF with rank $I-2$. Thus, a penalised spline can be seen from a Bayesian view as  placing an improper, zero-mean GRF prior with rank $I-2$ precision matrix on $\bs{\gamma}$. The RW2 model is an example of one such prior, and furthermore, for some spline representations, $\bs{S}$ can have exactly the same tri-diagonal form as $\bs{R}$ \citep{wood2006generalized}. 
%is The connection between a penalised spline and a RW2 prior can be seen as the Bayesian interpretation of the penalised likelihood is to place an improper, zero mean GRFs with a precision matrix of rank $n-2$ prior on $\bs{\beta}$, and a RW2 is a specific one of these priors \textcolor{red}{Difficulat sentence}. 

% on a regular lattice \citep{wood2006generalized, yue2012priors}. The impropriety in the prior is due to the penalty matrix $\bs{S}$ not being of full rank, meaning the precision matrix $\bs{Q}_{\lambda}$ is not invertible. The precision matrix is rank deficient by the dimension of the null space for the penalty matrix $\bs{S}$. A random walk of order $p$ prior is an improper Gaussian prior (with a Markov property) which has a rank deficiency of $p$. If we choose to use a RW2 prior, then $\bs{Q}_\lambda$ will be an improper GMRF with a rank deficiency of two. This is equivalent to using the cubic penalty function since $\bs{S}$ is rank deficient by two. Because of this, both the RW2 prior and the penalty are penalising deviations in linearity. 

Whilst both the penalised smoothing spline model and the RW2 prior model are imposing a penalty on the second derivative of the estimates, how each model performs this differs which causes slight differences in practical estimates. For example, the smoothing parameter of the penalised smoothing spline model can be estimated by cross validation \citep{wood2006generalized}, e.g. in the R-package \mgcv, whereas estimation of the equivalent smoothing parameter of the RW2 prior model is based upon the choice of the prior distribution \citep{rue2005gaussian}, and can be estimated for example in \inla. The data driven approach of the penalised smoothing spline model is attempting to find an `optimal’ smoothing parameter $\lambda$; whereas in the RW2 prior model, we specify a distribution for the smoothing parameter $\tau$, \textit{a priori}.

\section{Simulation study} \label{Section: Simulation Study}

The simulation study is motivated by the alcohol-related deaths from Section \ref{Section: data}. The shapes for the age, period, and cohort effects are adopted from a simulation study for Gaussian data from \citet{luo2016block}. To keep the shapes of the age, period, and cohort functions but make the responses representative of the rare alcohol-related deaths example, we included a scale and shift alongside the functions of \citet{luo2016block}. A similar alteration was performed in \citet{gascoigne2023penalized}.

We generated observations using single-year ages (from 10 to 84) and periods (2000 to 2020). We fixed 750,000 as the population at risk for each age-period combination to align with the population size in the alcohol and self-harm-related deaths example. To mimic the reality of data collection and dissemination, we generated data in single-year age-period combinations and then aggregated age into five-year groups, i.e., $10 - 14, 15 - 19, \dots, 80 - 84$. When modelling, we used the midpoint of the ages groups, i.e., $12.5, 17.5, \dots, 82.5$. We simulated $m = 1, \dots, M = 100$ data sets in this way. For each data set, we assess both the estimation and predictive capabilities. We fit the model for all years between 2000 and 2017 and forecast for years 2018 to 2020.

For the simulation study, we used three common spline basis functions discussed previously: CRS, BS, TPRS. As shown in Figure \ref{Fig: whyIncludePenalty}, the choice of the number of knots is less important when including a penalty. Consequently, we use 10, 10, and 12 for the number of age, period, and cohort knots, respectively, when defining their basis functions. We fit a RW2 model with three different PC prior specifications. For all specifications, we used $\alpha = 0.01$ and either $U = 1$, $U = 3$, or $U = 6$.

\subsection{Assessment criteria}

We assessed the models estimation and prediction performance using the Mean Absolute Error (MAE) and Mean Square Error (MSE) computed separately for the estimation and in-sample prediction. We defined the MAE and MSE as
\begin{equation*}
    \label{Eq: MAE and MSE}
    \text{MAE}_{ap} = \frac{1}{M} \sum_{m=1}^{M} \abs{\widehat{\eta}_{ap} - \eta_{ap}} \quad \text{and} \quad \text{MSE}_{ap} = \frac{1}{M} \sum_{m=1}^{M} \pa{\widehat{\eta}_{ap} - \eta_{ap}}^2
\end{equation*}
where for age $a$ and period $p$, $\widehat{\eta}_{ap}$ and $\eta_{ap}$ are the fitted and true log rates, respectively.

In addition to the MAE and MSE, we assessed the entire predictive distribution using the 95\% Interval Score (IS) \citep{gneiting2007strictly}. The IS is a scoring rule that transforms interval width and empirical coverage into a single score. Our estimated log rate for each age $a$ and period $p$ combination is associated with lower and upper uncertainty, $\bra{l_{ap}, u_{ap}}$, defined using the respective $\pa{1 - \alpha} \cdot 100 \%$ lower and upper predictive quantiles. See \ref{SubSec: implementation} for how these limits are calculated. The IS for $\alpha \in \pa{0,1}$ is defined 
\begin{equation*}
    \label{Eq: Interval Score}
    \text{IS}_{\alpha}\pa{\eta_{ap}} = \pa{u_{ap} - l_{ap}} + \frac{2}{\alpha}\pa{l_{ap} - \eta_{ap}}\II\bra{\eta_{ap} < l_{ap}} + \frac{2}{\alpha}\pa{\eta_{ap} - u_{ap}}\II\bra{\eta_{ap} > u_{ap}}
\end{equation*}
where $\II\bra{\cdot}$ is an indicator function that penalises how many data points, here $\eta_{ap}$, %$ (in our examples, these are the counts divided by the population) 
are outside the interval. The final score is defined by averaging over all IS's, $\text{IS}_{\alpha} = \Sigma_{ap} \text{IS}_{\alpha}\pa{\eta_{ap}}$. A lower IS is indicative of a better performing model.

\subsection{Implementation} \label{SubSec: implementation}

We fitted the penalised spline model with the Generalised Additive Model (GAM) framework, as implemented in the \mgcv \ package \citep{wood2006generalized}. This package offers a wide range of spline bases to represent smooth functions and their penalties. In \mgcv, the syntax to fit a penalised spline on the term \texttt{x} is \texttt{s(x, k, bs, fx = FALSE)}. The argument $k$ is for the number of knots used to define the basis function and \texttt{bs} defines the type of basis function. For the CRS, BS and, TPRS functions, \texttt{bs = `cr'}, \texttt{`bs'} and \texttt{`tp'}, respectively. The argument \texttt{fx = FALSE} (which is the default option) ensures a penalised (as oppose to un-penalised) smoothing spline is being fit. 

We fitted the RW2 model within a Bayesian hierarchical framework using Integrated Nested Laplace Approximation (INLA), as implemented in the \inla \ package \citep{rue2009approximate}. INLA provides accurate approximations of the marginal posterior distribution for all model parameters whilst avoiding the need for costly and time-consuming Markov-chain Monte Carlo (MCMC) sampling. In \inla, the syntax to fit a RW2 prior on the term \texttt{x} is \texttt{f(x, model = `rw2', hyper, ...)}. The argument \texttt{model = `rw2'} specifies we fitted a RW2 model. By fitting a RW2 model \inla \ will implicitly set the arguments \texttt{contr = TRUE, rankdef = 2} to constrain the model which, in the case of a RW2, is to constraint against an intercept and linear trend which automatically forces the rank deficiency of the model to be two. The argument \texttt{hyper} is where the hyper priors are specified. PC priors were specified by \texttt{list(prec = list(prior = `pc.prec', param = c($U$, $\alpha$)))} where we used $\alpha = 0.01$ and $U = 1, 3$ and $6$. 

For the results relating to a penalised spline model, the point estimate is the value found through maximising the penalised likelihood and the associated uncertainty is calculated by adding and subtracting the standard error for each estimate multiplied by $1.96$, the 97.5th percentile point of the normal distribution. The standard errors in \mgcv \ are based on the Bayesian posterior covariance matrix \citep[Section 4][;]{wood2006generalized}. For the results relating to a RW2 model, the point estimate is the median, i.e., 50\% quantile, of the distribution of interested and the associated uncertainty are the 2.5\% and 97.5\% quantiles. 

\sloppy We have provided all the relevant code and data in the online GitHub repository \url{https://github.com/connorgascoigne/Smoothing-for-APC-models-splines-and-random-processes}. For the simulation study, there is the code to generate the simulated data and run the analysis. In addition, we provide the simulated data we used. For the alcohol and self harm related deaths illustration, we provide the code to run the analysis as well as the data we used, which fall under a UK Open Government License (https://www.nomisweb.co.uk/home/copyright.asp). 

\subsection{Results}

\begin{figure}[!h]
    \centering
    \includegraphics[width=\linewidth]{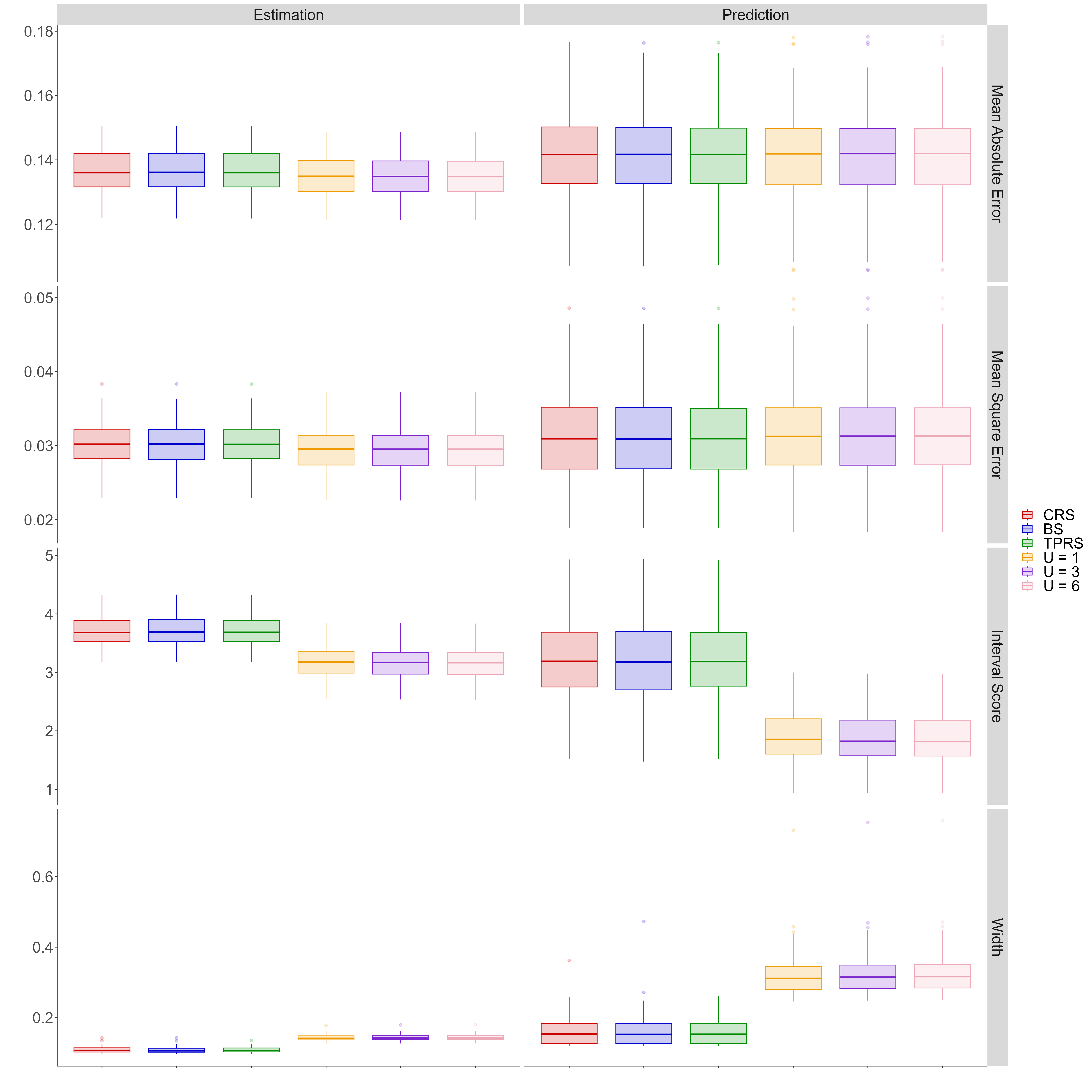}
    \caption{Boxplots for Mean Absolute Error, Mean Square Error, Interval Score and (uncertainty) width for the simulation study. The column facet are the scores for the estimated (left) and predicted (right) values, respectively. The row facets are for each of the scores listed above going from top-to-bottom, respectively. In each facet, the first three boxes are spline models defined with Cubic Regression Splines (CRS), B-Spline (BS) and Thin Plate Regression Spline (TPRS) basis. The last three boxes are Random Walk 2 models define using $U = 1, 2$ and $6$ in the Penalised Complexity priors.}
    \label{Fig: boxplots}
\end{figure}

Figure \ref{Fig: boxplots} shows the the results from the simulation study as a range of boxplots. The top-to-bottom rows are for the assessment criteria MAE, MSE, IS and 95\% (uncertainty) width. The left-to-right columns are for whether the model values from the model were either estimates or predictions. The red, blue and green colours are the results of the penalised spline models with a CRS, BS and TPRS basis specification, respectively. The yellow, purple and pink colours are the RW2 model results with $U = 1$, $3$ and $6$ used in the PC priors, respectively. 

First, we consider the results for estimation. For all scores, there is little difference within each model for the different specifications. Therefore, the results are robust to how the penalised spline and the RW2 are specified. For the MAE and MSE, the difference between the penalised splines and RW2 models is very small and negligible when considering the scale of the $y$-axis. For the IS, the RW2 models noticeably outperform the penalised splines, with the Interquartile Range of the boxplot for the RW2 models being below those of the penalised spline. When considering the widths, the RW2 models are larger than those of the penalised spline. As the IS penalises the 'true' value being outside of the uncertainty intervals, the RW2 models having a better IS than the penalised spline models indicates that the penalised spline uncertainty intervals are too narrow. %, hence do not capture the variation of the data as well as the RW2 models. 

Now, we consider the results for prediction. For all scores, the results are robust to the way each model is specified. The MAE and MSE only take the point prediction into account, leading to no noticeable difference between the results from the penalised spline or RW2 models. Considering the entire distribution of the log-rate, we find that the interval scores for the RW2 models are clearly lower than those of the penalised splines. The widths for the RW2 models are larger than those for the penalised splines. When comparing the difference between the IS and width results for the penalised splines and the RW2 models, this difference is larger in the predictions than in the estimates. Therefore, the narrow widths of the penalised splines become more detrimental to the model performance, with respect to the distributional scores, as predictions are subject to much more uncertainty than estimates, and the further forward the predictions, the more uncertainty there is.

%Whilst the MAE, MSE, IS and width results show the two approaches are similar, they are not exact like one would assume based on the theoretical link. 
Differences in the MAE and MSE are due to differences in how each approach estimates the smoothing parameter. The smoothing parameter for the penalised smoothing spline is found through cross-validation, whereas the smoothing parameter in the RW2 model is defined \textit{a priori} and updated from the data using Bayes Theorem. Differences in the IS and width are due to the way each measure quantifies uncertainty and defines the uncertainty interval. 

\section{Real data analysis: death due to alcohol and intentional self-harm}

Based on the results of the simulation study, the outcomes were robust to the specification of the model, i.e., the type of spline basis and the prior specification. Therefore, for the alcohol and self-harm-related deaths examples, we used the TPRS spline basis, the \mgcv\ default, and $U = 1$, the \inla\ recommendation, to specify the penalised spline and RW2 models, respectively. For the alcohol and self-harm-related deaths datasets, we used all ages in the years 2006 to 2017 to fit the models and assess the models' estimation (in-sample predictions) when compared to the real data. Using the models fit to the years 2006 to 2017, we predicted the years 2018 to 2021 and used this window to assess the models' (out-of-sample) predictions when compared to the real data.

Table \ref{Tab: Suicide data model fit Scores} shows the IS and width (distributional scores) for both outcomes and each model partitioned into estimation and prediction results. For both estimation and prediction on both outcomes, the RW2 model always has the lower IS score, suggesting a better-performing model. When considering the width, the RW2 model has a larger width, which indicates the RW2 models have a better distribution score as the penalised spline is not capturing the variation between the points as well; it produces a narrow interval. The RW2 outperforming the penalised spline is more pronounced when considering prediction. %Whilst the estimation scores are closer, the prediction scores are more in favour of the RW2 model as ther Bayesian implementation does a far betetr job accounting for uncertainty in the predictions (where uncertainty is larger).

\begin{table}[!h]
    \centering
    \caption{Model scores for the alcohol and self harm related death data. %The First two rows are for alcohol related deaths and the latter two rows are for self harm related deaths. The columns are split into scores for estimation and prediction.
    }
    \label{Tab: Suicide data model fit Scores}
    \begin{tabular}{c|c|cc|cc}
        \hline
        \multirow{2}{*}{\textbf{Dataset}} & \multirow{2}{*}{\textbf{Model Type}} & \multicolumn{2}{c|}{\textbf{Estimation} $\pa{\times10^{-2}}$} & \multicolumn{2}{c}{\textbf{Prediction} $\pa{\times10^{-2}}$} \\
         &  & \textbf{Interval Score} & \textbf{Width} & \textbf{Interval Score} & \textbf{Width} \\
         \hline
        \multirow{2}{*}{Alcohol} & Spline & 323.93 & 23.77 & 167.04 & 48.74 \\
         & RW2 & 275.22 & 27.16 & 120.44 & 104.03 \\
         \hline
        \multirow{2}{*}{Self harm} & Spline & 95.90 & 9.57 & 108.54 & 18.16 \\
         & RW2 & 76.51 & 10.97 & 71.97 & 36.04 \\
         \hline
    \end{tabular}
\end{table}

\begin{figure}[!h]
    \centering
    \begin{subfigure}{\textwidth}
        \centering
        \includegraphics[width=\linewidth]{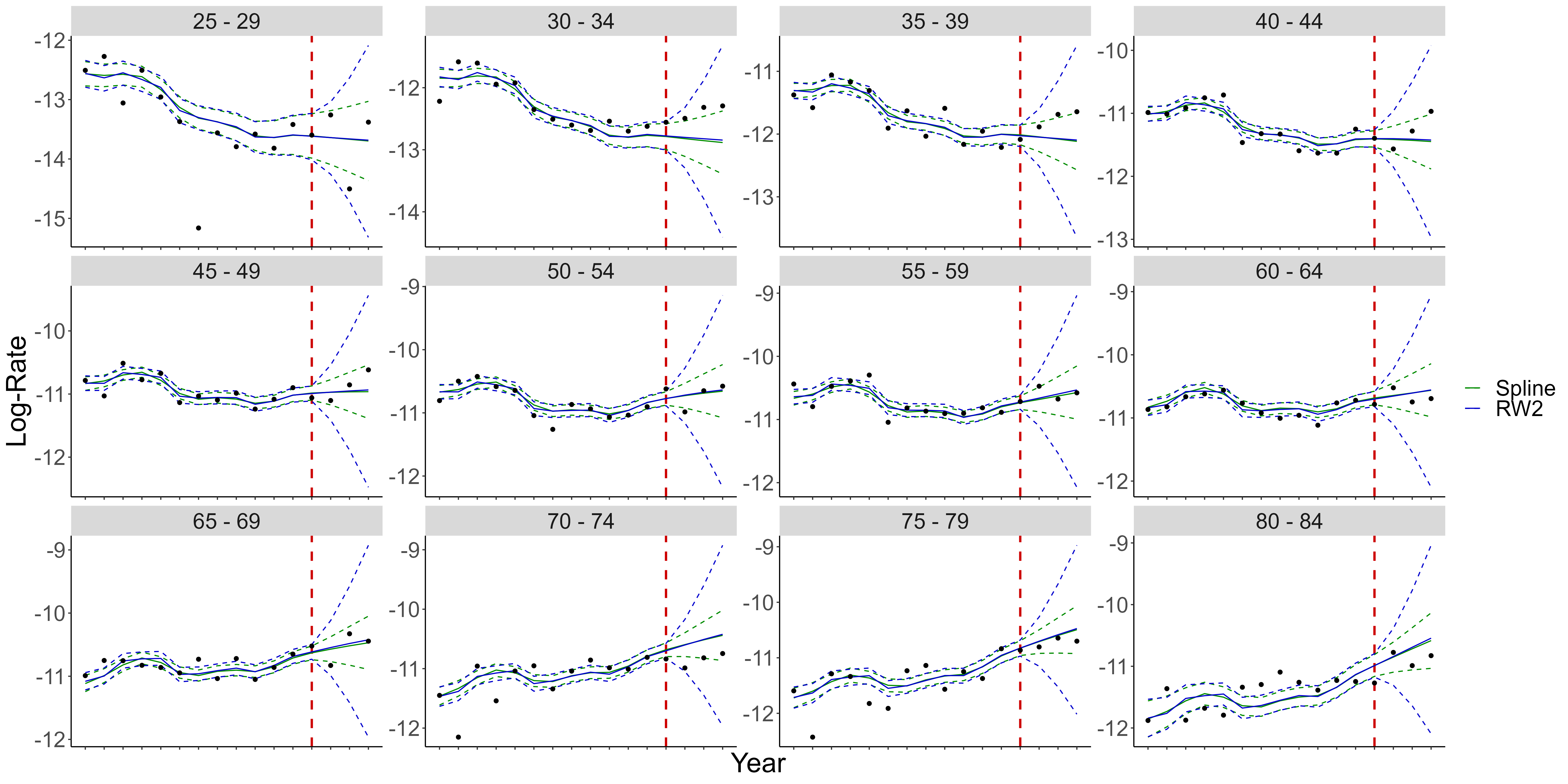}
        \caption{Alcohol related deaths}
        \label{Fig: alcoholPredictedLineplot}
    \end{subfigure}
    
    \begin{subfigure}{\textwidth}
        \centering
        \includegraphics[width=\linewidth]{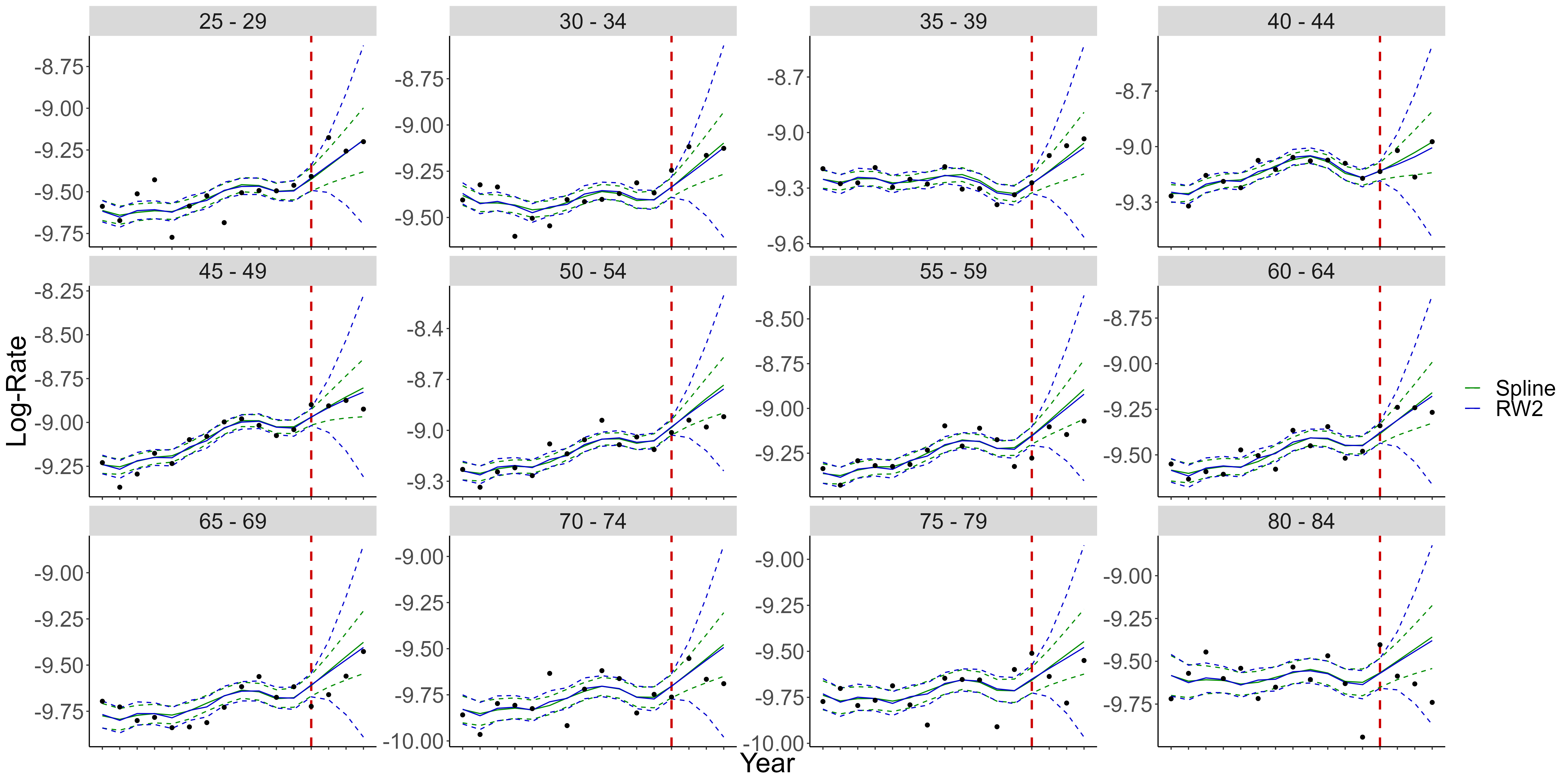}
        \caption{Self harm related deaths}
        \label{Fig: selfHarmPredictedLineplot}
    \end{subfigure}
    \caption{Estimated and predicted values of $\eta_{ap}$ for the spline and random walk 2 models for alcohol (a) and self harm (b) related suicides. In each subfigure, the facets are for each of the age groups increasing from left-to-right and top-to-bottom. The $y$-axis is the log rate and the $x$-axis is the year. The green and blues solid lines are the fitted valued for the spline and random walk 2 models, respectively. The dashed lines are their associated uncertainty levels. The black dots are the true values and the vertical red dotted line is where the estimation stops and prediction begins.}
    \label{Fig: PredictedLineplot}
\end{figure}

To demonstrate why the RW2 model produces a better IS for both estimation and prediction than the penalised spline model, we present the model estimates and predictions against the real data for both outcomes in Figure \ref{Fig: PredictedLineplot}. In Figure \ref{Fig: PredictedLineplot}, the estimated values (solid lines) for the penalised smoothing spline (green) and RW2 (blue) models, along with their associated lower and upper uncertainty intervals (dashed lines), are shown for each age group for alcohol (Figure \ref{Fig: alcoholPredictedLineplot}) and self-harm (Figure \ref{Fig: selfHarmPredictedLineplot}) related deaths. %To highlight why the RW2 model performs better for the IS, we superimposed 
The real data (black dots) are superimposed over the estimates. The combination of a better IS and a larger uncertainty width for the RW2 models can be seen if one of the black dots falls within the blue dashed lines but outside the green dashed lines. This is clearer to see when considering the predicted years (after the vertical red dashed line). For example, consider the last data point in facet 40 -- 44 for alcohol-related deaths, Figure \ref{Fig: alcoholPredictedLineplot}. This point falls within the RW2 uncertainty interval but not the penalised spline's uncertainty interval, and this would contribute to a better IS for the RW2 in comparison to the penalised spline. There are multiple more events such as this that contributed to the RW2's overall better performance in terms of the IS.

\section{Discussion}

In this article, we discuss the theoretical link between model fitting via smoothing splines and random processes. In the context of modelling health and demographic data, APC models are commonly used, with two distinct schools: those who use splines and those who use random processes. Using the theoretical link, we showed, through simulated and real data, that model fitting via penalised smoothing splines and random processes are comparable in the context of APC models.

For both the simulated and real data examples, we assessed model performance using a range of scores, which we partitioned into scores for estimation and prediction. APC models are often used for predictive purposes, as forecasts for the burden of future health concerns are an important goal for many policymakers. When considering the point estimates only, the use of penalised splines and RW2 models is interchangeable. This is shown by the MAE and MSE from the simulation study having little-to-no difference at all. Furthermore, a similar conclusion can be made when considering the RW2 versus penalised spline plot line in the alcohol and self-harm-related deaths example. However, when one wishes to include uncertainty in the results, the results become different. The narrower confidence intervals of the smoothing spline approach are reflected in a model that does not capture variation in the data as well as a model fit using RW2 models in a Bayesian paradigm. The 'Interval Score' of \citet{gneiting2007strictly}, which defines a score that balances the width of the uncertainty interval and whether or not the observation falls within, highlights this. The inclusion of uncertainty in estimates is vital for policymakers as it allows them to base any future policy on the worst, middle, and best-case scenarios.

While the two methods are equivalent due to their theoretical link, the differences can be attributed to how each method approaches smoothing in their respective software. For smoothing splines in \mgcv, the smoothing parameter is estimated from the data using cross-validation. For random processes in \inla, the smoothing parameter is defined \textit{a priori} and is updated from the data. Generally, this needs to be done carefully as Bayesian methods can be prone to over/under-smoothing for a poor choice of prior. Given the similarities between the methods, if a researcher naively chooses one implementation over the other, our results show they should not worry about the analysis producing very different results. However, if the researcher were to make a nuanced choice to better account for uncertainty, they would choose to use random processes (implemented via the Bayesian paradigm). It is worth noting that smoothing splines can also be fit in a fully Bayesian workflow using \inla \ \citep{bauer2016bayesian} or with MCMC using packages such as \texttt{brms} \citep{burkner2018advanced}, though, to the best of our knowledge, these have not been considered in an APC context.

When modelling health and demographic rates, there are alternatives to the APC model that can be used. To avoid the co-linearity between the three temporal effects, an age-period interaction model or a model where the cohort is replaced with a proxy, as discussed by \citet{clayton1987b}. Both methods aim to replace the cohort with an equivalent term that is not linearly dependent on age and period. However, as we focus on identifiable APC models, we did not consider these methods here. Another class of models commonly used for mortality modelling is the Lee-Carter model \citep{lee1992modeling}. We expect that our findings also extend to smoothing within this model, however, this is left for future work.

%\begin{equation*}
 %   \label{Eq: Lee-Carter Model}
 %   \lambda_{ap} = \alpha_{a} + \beta_a \kappa_{t}
%\end{equation*}
%where $\alpha_a$ is the general shape of mortality by age, $\kappa_p$ is a time index of the general level of mortality, and $\beta_a$ describes the extent to which mortality at age $a$ changes given the overall temporal change in the general level of mortality. Since we are using \inla, we require linearity in the linear predictor which is not the case for Lee-Carter models as the term $\beta_a \kappa_{t}$ is multiplicative.

In conclusion, APC models are widely used tools for modelling health and demographic data, within which smoothing is a vital piece of the puzzle. Smoothing for APC models is implemented using two main schools: frequentist (penalised splines) and Bayesian (random processes) methods. While these two methods are interchangeable and produce similar results, we have shown that if a researcher wants to better capture the uncertainty of their data and provide a more complete inference, then Bayesian models can prove beneficial. With software such as \inla, model fitting via random processes has become increasingly more accessible, with the need for specialist knowledge reducing thanks to intuitive and reliable default choices within the software.

\clearpage